\documentclass[preprint2]{aastex}

\newcommand{\etal}{\mbox{et~al.}}

\def\deg      {{\ifmmode^\circ\else$^\circ$\fi}} 

\shorttitle{COSMOS: Strong Lens Systems}
\shortauthors{Faure et al.}

\begin{document}

\title{First catalog of strong lens candidates in the COSMOS field}


  \author{
Cecile Faure\altaffilmark{1,2},
Jean-Paul Kneib\altaffilmark{3,4},
Giovanni Covone\altaffilmark{3,5},
Lidia Tasca\altaffilmark{3},
Alexie Leauthaud\altaffilmark{3},
Peter Capak\altaffilmark{4},
Knud Jahnke\altaffilmark{11},
Vernesa Smolcic\altaffilmark{11},
Sylvain de la Torre\altaffilmark{3},
Richard Ellis\altaffilmark{4},
Alexis Finoguenov\altaffilmark{10},
Anton Koekemoer\altaffilmark{7},
Oliver Le Fevre \altaffilmark{3},
Richard Massey\altaffilmark{4},
Yannick Mellier\altaffilmark{8},
Alexandre Refregier\altaffilmark{9},
Jason Rhodes\altaffilmark{4},
Nick Scoville\altaffilmark{4},
Eva  Schinnerer\altaffilmark{11},
James Taylor\altaffilmark{4,12},
Ludovic Van Waerbeke\altaffilmark{6},
Jakob Walcher\altaffilmark{3}
}

%
\altaffiltext{1}{Astronomisches Rechen-Institut, Zentrum f\"ur Astronomie der Universit\"at Heidelberg, M\"onchhofstr.
12-14, 69120 Heidelberg, Germany}
\altaffiltext{2}{Departamento de Astronom\'{i}a, Pontificia
Universidad Cat\'olica, Av Vicu\~na Mackenna 4860, 782-0486 Macul,
Santiago, Chile}
\altaffiltext{3}{Laboratoire d'Astrophysique de Marseille, UMR6110, CNRS-Universit\'e de Provence, BP8, F-13376 Marseille cedex 12, France}
\altaffiltext{4}{California Institute of Technology, MC 105-24, 1200
East
California Boulevard, Pasadena, CA 91125}
\altaffiltext{5}{INAF, Osservatorio Astronomico di Capodimonte,
Naples, Italy}
\altaffiltext{6}{Department of Physics \& Astronomy, University of
British Columbia, 6224 Agricultural Road, Vancouver, B.C. V6T 1Z1,
Canada}
\altaffiltext{7}{Space Telescope Science Institute, 3700 San Martin
Drive, Baltimore, MD 21218, U.S.A.}
\altaffiltext{8}{Institut d'Astrophysique de Paris, UMR 7095, 98 bis
Boulevard Arago, 75014 Paris, France}
\altaffiltext{9}{Service d'Astrophysique, CEA/Saclay, 91191 Gif-sur-
Yvette, France}
\altaffiltext{10} {Max-Planck-Institut f\"ur extraterrestrische
Physik, Giessenbachstrasse, 85748, Garching, Germany}
\altaffiltext{11}{Max-Planck-Institut f\"ur
Astronomie, K\"onigstuhl 17, 69117 Heidelberg, Germany}
\altaffiltext{12}{Department of Physics and Astronomy, 
 University of Waterloo, Waterloo, Ontario N2L 3G1, Canada}

\begin{abstract}
  We present the first catalog of 67 strong galaxy-galaxy lens candidates
  discovered in the 1.64 square degree {\it Hubble Space Telescope}
  COSMOS survey. Twenty of these systems display multiple images or
  strongly curved large arcs. Our initial search is performed by
  visual inspection of the data and is restricted, for practical
  considerations, to massive early-type lens galaxies with arcs found
  at radii smaller than $\sim$5\arcsec. Simple mass models are
  constructed for the best lens candidates and our results are compared to
  the strong lensing catalogs of the SLACS survey and the CASTLES
  database. These new strong
  galaxy-galaxy lensing systems constitute a valuable sample to study the
  mass distribution of early-type galaxies and their associated dark
  matter halos. We further expect this sample to play an important
  role in the testing of software algorithms designed to automatically
  search for strong gravitational lenses. From our analysis a robust lower limit is derived for the expected
  occurrence of strong galaxy-galaxy systems in current and future
  space-based wide-field imaging surveys. We expect that such surveys
  should uncover a large number of strong lensing systems (more than
  10 systems per square degree), which will allow for a detailed
  statistical analysis of galaxy properties and will likely lead to
  constraints on models of gravitational structure formation and
  cosmology. The sample of strong lenses is available at this address: http://cosmosstronglensing.uni-hd.de/  
\end{abstract}



\section{Introduction}

Gravitational lensing is one of the most promising tools in modern
astrophysics for probing the matter content of the universe directly, and
thereby constraining cosmological models.   Applications of strong 
 gravitational lensing include
measurements of the Hubble constant using lensed quasars with
  time-delays (Refsdal 1964; {\it e.g.} of time delay measurements:
  Barkana 1997; Cohen et al.~2000; Burud et al.~2000; Burud et
  al. 2002~a,b; Fassnacht et al.~2002; Hjorth et al.~2002,
  Ofek \& Maoz 2003; Jakobsson et al.~2005; Kochanek et al.~2006;
  Fohlmeister et al.~2006; Vuissoz et al.~2007) and the determination
  of galaxy and cluster mass distributions using multiple images ({\it
    e.g.} Guzik \& Seljak 2002, Mandelbaum et al.~2006, Read et
  al.~2007). Moreover, homogeneous samples of strong lenses allow
  for a statistical determination of the properties of lensing
  galaxies ({\it e.g.}  Kochanek 1996; Rusin \& Ma 2001; Chae, Mao \&
  Kang 2006, Chae 2007) and samples of strong lenses selected based on
  source properties are used to test cosmological models ({\it
    e.g.}  Fukugita \& Turner 1991; Wambsganss et al.~2005; Moeller,
  Kitzbichler \& Natarajan~2006; Hilbert et al.~2007).

The giant arcs that are observed around massive clusters of galaxies
are among the most spectacular lensing phenomena ({\it e.g.}
  Gladders et al.~2002 searched giant arcs in the Red-Sequence Cluster
  Survey (RCS); Scarpine et al.~2006, Hennawi et al.~2006, Estrada et
  al.~2007 in the Sloan Digital Survey (SDSS)).  These cases
are relatively rare because each square degree of sky contains only
about one cluster of sufficient mass to produce a giant arc. Much
more common are the strong lensing events found around massive
early-type galaxies. The first studies of galaxy-galaxy strong lensing
systems targeted distant, {\it bright} lensed sources,
including quasars (ESO Hamburg survey: Wisotzki et al.~1996;
CfA-Arizona-ST-LEns-Survey: Peng et al.~1997, Mu\~noz et al.~1998) and
radio sources (Jodrell Bank-VLA Astrometric survey: Patnaik et
al.~1992, Browne et al.~1998; Cosmic Lens All Sky Survey: Jackson et
al.~1995, Myers et al.~1995). The identification of multiple images in
these systems is generally unambiguous and does not require very deep
observations; but requires imaging over a large fraction of the sky.

The advent of deep and wide optical imaging combined with large
spectroscopic surveys has greatly broadened the opportunities and
means by which strong galaxy-galaxy lensing events can be
found. Indeed, since we expect that $\sim$0.1\% of massive galaxies
(the luminous elliptical galaxies) strongly distort and magnify a
galaxy source at z$>$1 (Miralda-Escud\'e \& L\'ehar 1992), new wide field surveys
are likely to discover many hundreds of fainter galaxy-galaxy strong
lensing systems. Spectroscopic identification of two galaxies at
different redshifts in the same Sloan Digital Sky Survey fiber has
proved to be a successful method in the Sloan Lens ACS survey (SLACS:
Bolton et al.~2006). Deep, wide-field imaging surveys (especially
those utilizing the exquisite image resolution available from space)
provide another avenue for locating strong galaxy-galaxy lenses.  This
was first attempted in the {\it Hubble Space Telescope} Medium Deep
Survey (Griffith et al.~1994, Ratnatunga et al.~1999), and more recently in the Great Observatories Origins Deep
Survey field (GOODS; Dickinson et al.~2001, Fassnacht et
al.~2004) or in the AEGIS survey (Moustakas et
al.~2006). Automated software for identifying strong lenses in {\it HST}
image archives are currently under development (Marshall et al.~2005,
Moustakas et al.~2006) and appear to be very promising. Searches for
strong lens systems are also underway from ground-based wide-field
surveys. In particular the SL2S initiative (Cabanac et al.~2007) was
able to find a few lenses per square degree in the CFHT-LS wide survey. However, ground based imaging surveys are
necessarily limited to finding systems with large deflection angles because
of the seeing limitation compared to space-based observations.

The 1.64 square degree {\it Hubble Space Telescope} COSMOS survey
(Scoville et al.~2007) provides an excellent opportunity to locate and
study a large number of strong galaxy-galaxy lensing systems. It
includes the largest contiguous high-resolution astronomical imaging
survey ever performed from space. Although observations with the {\it
  HST} Advanced Camera for Surveys (ACS) have only been obtained in a
single band (F814W filter), multi-wavelength coverage of this
equatorial field is provided in the optical band by deep multicolor imaging from
Suprime on the {\it Subaru} telescope and Megacam at {\it CFHT}.

In this paper, we present the first systematic attempt to locate
strong galaxy-galaxy lensing systems in the COSMOS field. The data-set
used for our investigation is described in $\S$ \ref{data} and the
techniques used to uncover strong galaxy-galaxy lensing candidates are
discussed in $\S$ \ref{seccriteria}. Our findings are presented in
$\S$ \ref{candidates}. The lens potential modeling scheme is discussed
in $\S$ \ref{model} and we compare our lens sample to other strong
lens catalogs in $\S$ \ref{seccomparison}. Finally, the results are
discussed in $\S$ \ref{secdiscussion}. Throughout this paper, galaxy
magnitudes are quoted in the AB system and cosmological distances are
calculated assuming a $\Lambda$CDM model with parameters H$_0$ =
73~km~s$^{-1}$~Mpc$^{-1}$, $\Omega_m$=0.3, and $\Omega_\lambda$=0.7.

\section{Data}\label{data}
\begin{figure}
\begin{center}
\includegraphics[width=8cm]{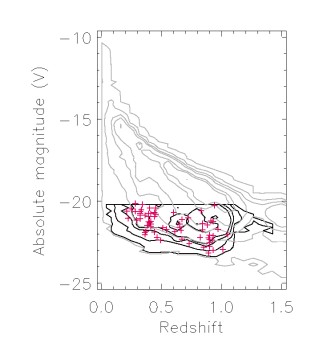}
\caption{\label{mvz}  The galaxies of the COSMOS field: V$_j$-band
    absolute magnitude versus photometric redshift. The grey solid line
    contours show isodensity numbers of all the galaxies in the entire
    photometric catalog (levels: 1, 100, 300, 500, 1000, 1200, 1600
    galaxies). The black solid line contours draw the isodensity numbers of
    galaxies in the `P'-catalog (levels: 1, 10, 30, 50, 100, 140
    galaxies). The red crosses point the lensing galaxy candidates described in Sect. \ref{candidates}. }
\end{center}
\end{figure}

To conduct our strong galaxy-galaxy lens search we take advantage of
three key data-sets of the COSMOS survey: the ACS/{\it HST} high resolution
imaging, the Suprime/{\it Subaru} imaging, and the Megacam/{\it CFHT}
multicolor imaging.

\subsection{HST data}

We have analyzed the full ACS field of view of the COSMOS field which
corresponds to 1.64 square degree observed during cycle 12 and 13. We
use data release v1.2 of the COSMOS ACS data which can be accessed
through the dedicated database at
IRSA\footnote{http://irsa.ipac.caltech.edu/data/COSMOS/images/}.
Individual exposures were combined using the MultiDrizzle software
(Koekemoer et al. 2007) onto an output grid of pixel size
0.03\arcsec. The limiting magnitude is about I$_{F814W}=26.5$~mag. A
detailed description of the data is available in Scoville et
al.~(2007), and a detailed description of the ACS-selected galaxy
catalog can be found in Leauthaud et al.~(2007).

\subsection{Subaru and CFHT data}

Multicolor imaging of the COSMOS field was obtained from the {\it Subaru}
telescope with the Suprime camera (Miyazaki et al.~2002) in $B_j, V_j,
g+, r+, i+, z+$ bands, and from the {\it CFHT} with the Megacam camera
(Boulade et al.~2003) in the $u*$ and $i*$ bands. In these data, the
seeing is generally better than 1\arcsec.  
 A
photometric catalog was derived from a combination of these data,
based on detections in the {\it Subaru} $i+$ band (Capak et al.~2007).

\subsection{Photometric redshifts}

The publicly available {\it{Le Phare}} photometric redshift estimation code (Ilbert et
al.~2006) has been used to measure the redshifts of 278\,526 galaxies
with $I_{F814W}<25$~mag. Details concerning the multi-wavelength
photometry can be found in Mobasher et al. (2007). The $I_{F814W}<25$~mag
limit ensures a good accuracy of both the photometric redshift
estimation and the fit to the spectral energy distribution (SED). The
ground-based photometric zero-points were calibrated (and the SED
templates adjusted accordingly) using 1095 spectroscopic redshifts
from the zCOSMOS Survey (Lilly et al.~2006). Using 8 bands, this
method achieves a photometric redshift recovery accuracy of
$\sigma_{\Delta z}/ (1+z_s) = 0.031$ with $\eta = 1.0 \%$ of
catastrophic errors, defined as $\Delta z /(1+z_s) >0.15$ (further
details can be found in Mobasher et al.~2007).


\section{Methodology}\label{seccriteria}

A four-step procedure is employed in order to identify strong
galaxy-galaxy lens systems. In order, the four steps are the
following:

\begin{enumerate}
\item Select a list of potential lenses from the photometric redshift
  catalog (hereafter `P' catalog\footnote{This catalog can be found at
    http://ari.uni-heidelberg.de/mitarbeiter/cfaure/downloads.html/Pcatalog.cat}).
\item Visually inspect the ACS images of all galaxies in the `P'
  catalog to produce a catalog of potential strong galaxy-galaxy lens
  systems (hereafter `E' catalog \footnote{This catalog can be found
    at
    http://www.ari.uni-heidelberg.de/mitarbeiter/cfaure/downloads.html/Ecatalog.cat}. Only 302 galaxies are in this catalog, the 35 others are missing from our personal archives.).
\item Investigate the `E' catalog using multicolor images to check for
  color differences between the main galaxy and the potentially lensed
  object.
\item Subtract a galaxy surface brightness model of the foreground
  galaxy to determine the morphology and lensing configuration of the
  potentially lensed background galaxy.
\end{enumerate}

In our methodology, systems that pass steps both 3 and 4 qualify as
strong galaxy-galaxy lens candidates. We will now describe each step
in more detail.

\subsection{The potential lens catalog}
\begin{figure}
\begin{center}
\includegraphics[width=7cm]{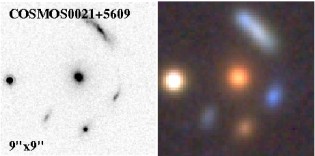}
\includegraphics[width=7cm]{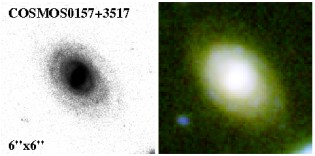}
\caption{\label{bad} Examples of strong lens candidates ruled out by
the pseudo-color image test. North is to the top of the figure and East is to the
left. The left stamp images are the ACS I-band images of the systems.
The right stamp  images are the color images.}
\end{center}
\end{figure}

The ACS-based photometric redshift catalog limited to $F814W=25$~mag
contains 278\,819 entries: it is unrealistic to inspect visually all
galaxies in this catalog. Furthermore, this would be inefficient,
because we know that a galaxy needs to be massive and not too distant
to efficiently distort and strongly magnify a more distant galaxy. We
therefore limit our search to intrinsically bright galaxies at
moderate redshifts. Furthermore, for this first investigation of the
COSMOS field, we only look at early-type galaxies (classified
spectroscopically by the photometric redshift code), which generally
have a simple, smooth surface brightness distribution. Once again,
we emphasize that the main reason for applying these selection
criteria to the photometric redshift catalog is to reduce the sample
of galaxies that need to be visually inspected. It is highly possible
that additional strong lens systems will be found around galaxies
that do not belong to our search sample. The number of strong lensing
systems that we find is thus a lower limit to the total number of
strong lensing systems in the COSMOS field.

To search for potential lenses, we select galaxies according to the
following criteria:

\begin{itemize}
\item Their photometric redshift must be in the range $0.2 \leq
  z_{phot} \leq 1.0$. The lower redshift bound of $z=0.2$ is chosen to
  ensure that galaxies will indeed produce lensing events and the
  upper bound is essentially motivated to ensure a good accuracy on
  the photometric redshift given that no deep infrared data are yet
  available for the COSMOS field. The galaxy is retained if the error
    bars at 68\% confidence level overlap, even partially, the redshift
    range [0.2;1.0].
\item Their luminosity must be $M_V < -20$~mag, on the assumption that the brightest galaxies are also likely to be the most
  massive ones with the largest cross-section for lensing.
\item We limit our search to early-type galaxies that have
  been spectrally classified as such by the SED fitting of our
  photometric redshift algorithm. The motivation for this criterion is
  to focus on massive systems with a simple light distribution that is
  easy to fit and to subtract.
\end{itemize}

We obtain a final `P' catalog of 9452 bright elliptical galaxies
(see Fig.~\ref{mvz}). Further strong lensing searches using automated
software will certainly relax these criteria.

During the course of our visual inspection, we serendipitously found
some additional strong lens systems in which the lensing galaxy did
not comply with the above limits. Although these constitute an
inhomogeneous sample, we add them to the list of strong lensing
systems for completeness. Those systems are identified in Tables
\ref{photom}, and \ref{photom1}.

\subsection{Visual inspection of the ACS images}\label{eyeballsect}

We inspect visually a postage-stamp 10\arcsec$\times$10\arcsec\, ACS
image surrounding all galaxies in the `P' catalog, looking for lensed
features (tangential arcs or multiple images). The search box size
limits us to r$_{arc}\lesssim 5$\arcsec.  Consequently, this puts an
upper limit on the mass scale for the lens, which excludes clusters
as a possible main deflector. This inspection was performed by  5 of the
co-authors using the {\sc
ds9} FITS image visualization tool. 

Our strategy reduced the 9452 galaxies in the `P' catalog
to an `E' catalog of 337 candidates, with intentionally loose
selection criteria to avoid excluding any real lenses at this stage.

\subsection{Pseudo-color images}

A color image is always useful to verify potential lens
candidates. Indeed, gravitational lensing is achromatic, thus a
gravitational arc or multiple images systems should have a consistent
color across the image(s). Furthermore, the color of a lensed galaxy
will only rarely have the same color as the main lensing galaxy. This
color test hence provides a good way to discriminate false strong
lensing cases.  However, with our particular combination of ground-
and space-based data (especially when the seeing approaches
1$\arcsec$), the color test is only useful for arcs more distant than
$\sim$0.5\arcsec\, from the lensing galaxy, because of strong
contamination from the wings of the lensing galaxy.
For the 337 systems in the `E' catalog, we produced pseudo-color
images made from the ``best PSF'' stacks of the $B$, $r$ and $i$
{\it Subaru} images, using the RGB feature in the {\sc ds9} software, as well as 
ground-based images sharpened with the ACS images.
The color images use the ACS F814W data as an illumination map and
the {\it Subaru} $B_J$, $r^+$, and $z^+$ images as a color map. To achieve
this each {\it Subaru} image is divided by the average of the three {\it Subaru}
images, then multiplied by the ACS F814W image. This preserves the
flux ratio between images, while replacing the overall illumination
pattern with the F814W data. Each image is then smoothed by a 1 pixel
Gaussian to reduce the noise and divided by $\lambda^2$ to enhance the
color difference between star forming and passive galaxies. The
processed $B_J$, $r^+$, and $z^+$ images are then assigned to the
blue, green, and red channels respectively. The resulting images have
the high spatial resolution of the ACS imaging, but color gradients
at ground based resolution.
 Of the 337 candidates, 67 have plausible colors for strong lensing systems. The color images are displayed in
Figs. \ref{figure11} to \ref{double}. For illustrative purposes, two objects that were rejected at this
stage are shown in Fig.~\ref{bad}. The system displayed in the top
panel of Fig.~\ref{bad} shows two arc-like features in the ACS image: a
giant arc located North-West of the main galaxy, and its candidate
counterpart, located to the South-West. The pseudo-color image
reveals that the ``arcs'' have very different colors, indicating that
they are not images of the same source: neither are likely to be
distorted images of background galaxies. Pronounced, arc-like features
are also observed in the ACS image shown in the bottom panel of
Fig.~\ref{bad}, North-East of the main galaxy bulge. With the
single-color ACS data-set only, it would not be clear whether these
features were part of the galaxy's spiral arms, or distorted images of
a background object. The pseudo-color image shows blue knots of star
formation all around the disc of this galaxy, suggesting that the
arc-like features are most probably genuine regions of the main
galaxy.

\subsection{The lensing galaxy fit and subtraction}\label{fitlight}
Fitting and extracting the foreground lens galaxy light from the
images reveals the shape of the more distant, lensed object. For this
work, a robust 2-dimensional fit to the galaxy surface brightness
distribution is derived using $GIM2D$ (Simard 1998, Marleau \&
Simard 1998).  We adopt a S\`{e}rsic bulge plus
exponential disc parametrisation to describe the two-dimensional
surface brightness distribution of the lensing galaxy light
profile. The S\'ersic law is parametrised by means of the total flux
in the bulge, the S\'ersic index $n$, the bulge ellipticity ($\epsilon
= 1-b/a$), the position angle of the bulge and the effective
radius of the bulge $R_{eff}$. The exponential profile depends on the
photometric disk total flux, the disk scale-length, the disk
position angle and the disk inclination. The $GIM2D$
software finds the best fitting values for all of these parameters,
and provides them as an output of the decomposition process.

In addition to the structural parameters for each object, we calculate
a model image of each galaxy, which is convolved with a point spread
function before comparison with real data. An image of the residuals is
obtained by the subtraction of the model from the science image.  We
present the result of the galaxy fits in Tables \ref{photom}, \ref{photom1} and in Figs.~\ref{figure11} to \ref{figure7}.
In most cases, the arc candidates appear clearly after the galaxy
model is subtracted (panel 3 in those figures).

In 10 cases, the modeling and subtraction of the lens galaxy proves
difficult. The problems encountered include irregular
galaxy morphology and compact galaxy size. We display the ACS and
color images of these systems in Fig.~\ref{double}. Additionally, the
system COSMOS~5737+3424 was discovered serendipitously in a part of
the ground-based data-set not covered by the ACS images. We present
just its pseudo-color image in Fig.~\ref{double}.

\section {Results of the strong lens  search}\label{candidates}



\begin{figure*}
\begin{center}

\includegraphics[width=14cm]{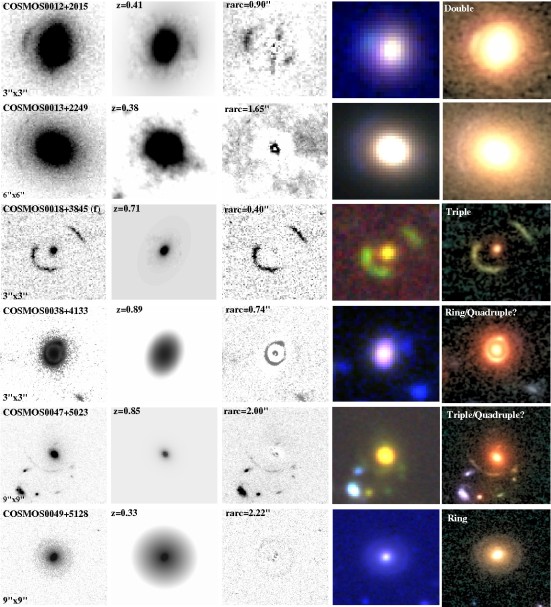}

\caption{\label{figure11} The first strong lensing candidates in the
COSMOS field: the best systems. From left to right: the first
panel is the ACS
I-band image of the lens. A letter (f) close to the name of an object
indicates that
it was found fortuitously. The second panel is the Gim2D model of the
lensing galaxy. The third panel is a subtraction of the two first
images. The fourth panel is the color image of the systems from
$B$,$R$,$I$
ground based images.  The fifth panel is the color image
sharpened with the
ACS-$F814w$ images. The label on the left top corner of panel five
indicates
the image multiplicity. When two multiplicities are given and are  separated by a slash, the second multiplicity is possible but difficult to confirm with the
present dataset.   North is to the top of the image and East is to the left. The photometric redshift of the lensing galaxy is displayed in
the second panel (error bars are given in Table \ref{photom}) and the
radius of the arc is displayed in the third panel.}
\end{center}
\end{figure*}

\begin{figure*}
\begin{center}
\includegraphics[width=14cm]{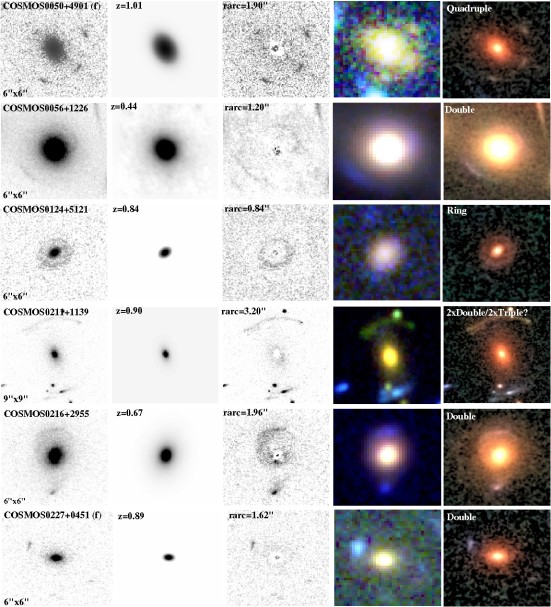}
\caption{\label{figure22} Continue Fig.~\ref{figure11}. }
\end{center}
\end{figure*}

\begin{figure*}
\begin{center}
\includegraphics[width=14cm]{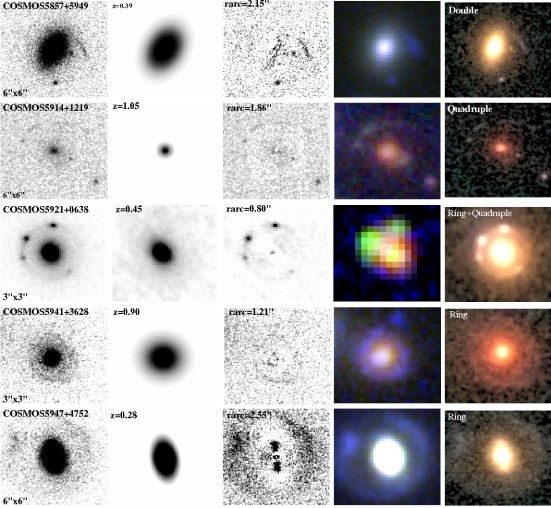}
\caption{\label{figure33} Continue Fig.~\ref{figure11}}
\end{center}
\end{figure*}

\begin{figure*}
\begin{center}
\includegraphics[width=14cm]{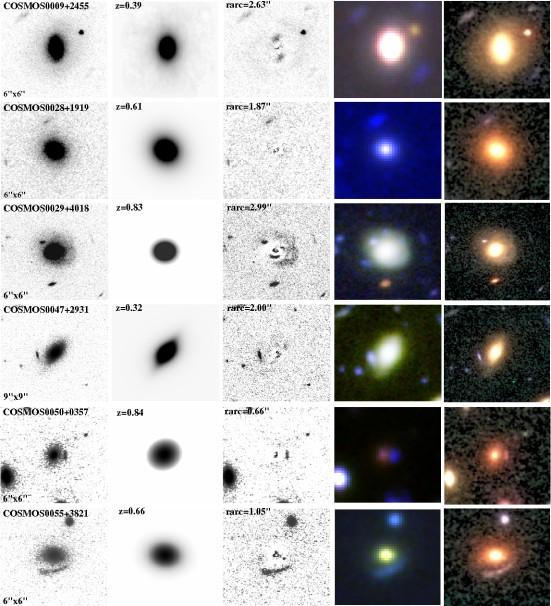}
\caption{\label{toto1} The first strong lensing candidates in the COSMOS
   field: the single arclet systems. From left to right:
   the first panel is the ACS I-band image of the lens. A letter
   (f) close to the name of an object indicates that it was found
   fortuitously. The second
   panel is the Gim2D model of the lensing galaxy. The third
   panel is the difference between the two first images. The fourth
panel is the color image of the systems from $B$,$R$,$I$
ground based images.  The fifth panel is the color image
sharpened with the
ACS-$F814w$ images. North is to the top of the image and East is to the left. The
   photometric redshift of the lensing galaxy is displayed in the
   second panel (error bars are given in Table \ref{photom1}) and the
   radius of the arc is displayed in the third panel. }
\end{center}
\end{figure*}

\begin{figure*}
\begin{center}
\includegraphics[width=14cm]{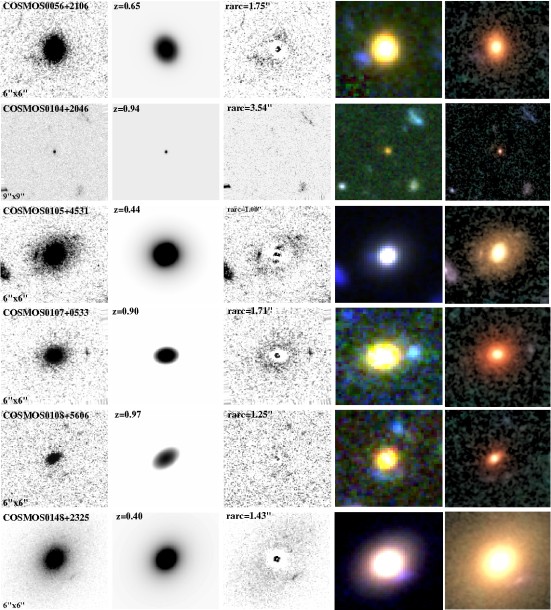}
\caption{\label{figure2}  Continue Fig.~\ref{toto1}. }
\end{center}
\end{figure*}

\begin{figure*}
\begin{center}
\includegraphics[width=14cm]{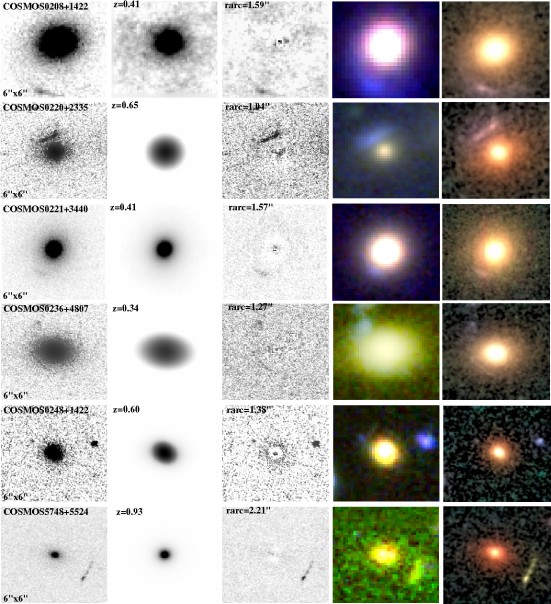}
\caption{\label{figure3}  Continue Fig.~\ref{toto1}. }
\end{center}
\end{figure*}

\begin{figure*}
\begin{center}
\includegraphics[width=14cm]{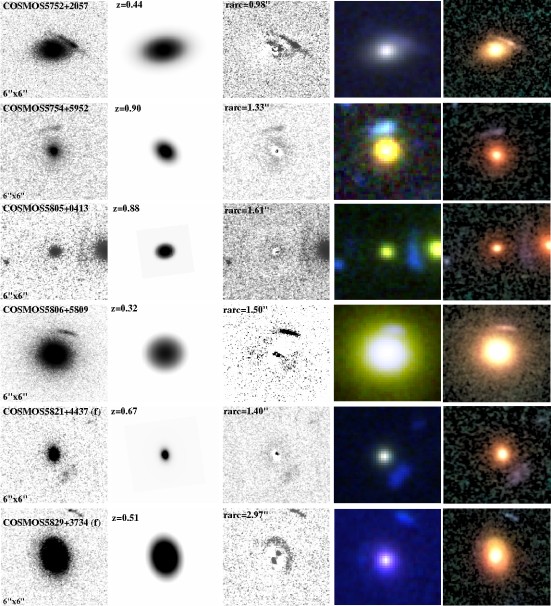}
\caption{\label{figure4} Continue Fig.~\ref{toto1}.  }
\end{center}
\end{figure*}

\begin{figure*}
\begin{center}
\includegraphics[width=14cm]{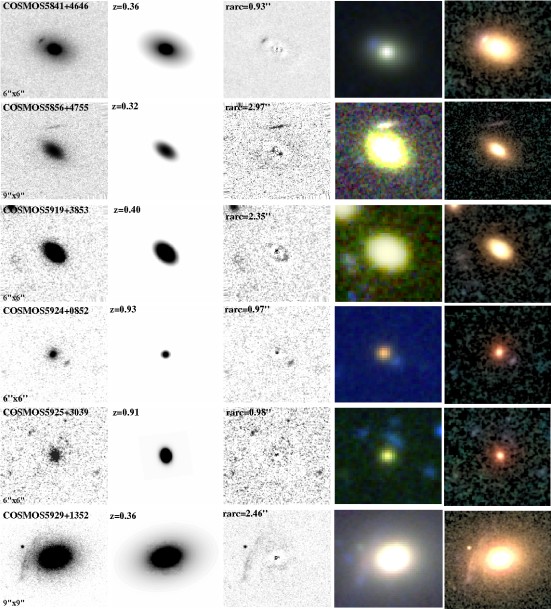}
\caption{\label{figure5} Continue Fig.~\ref{toto1}.  }
\end{center}
\end{figure*}

\begin{figure*}
\begin{center}
\includegraphics[width=14cm]{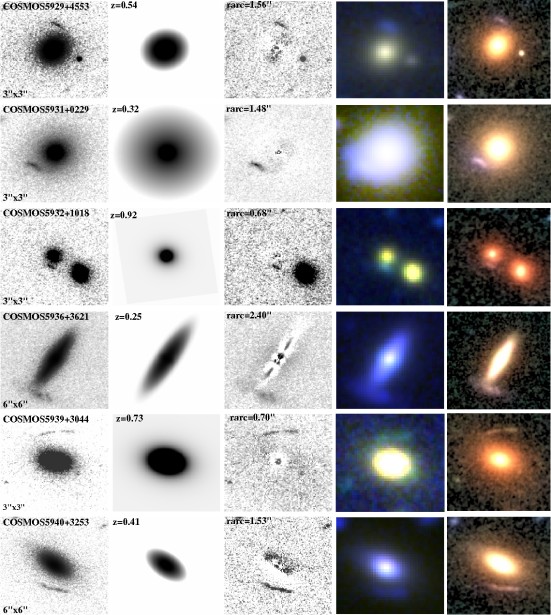}
\caption{\label{figure6} Continue Fig.~\ref{toto1}. }
\end{center}
\end{figure*}

\begin{figure*}
\begin{center}
\includegraphics[width=14cm]{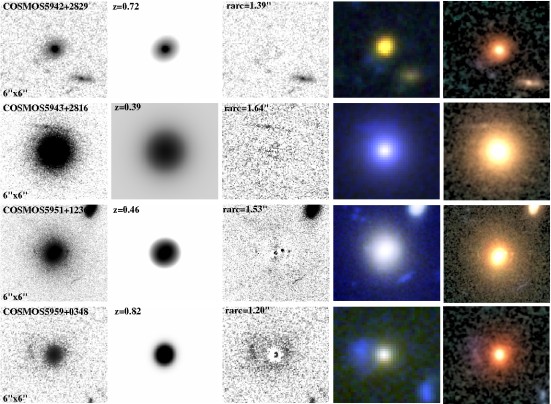}
\caption{\label{figure7} Continue Fig.~\ref{toto1}.  }
\end{center}
\end{figure*}

\begin{figure*}
\begin{center}
\includegraphics[width=14cm]{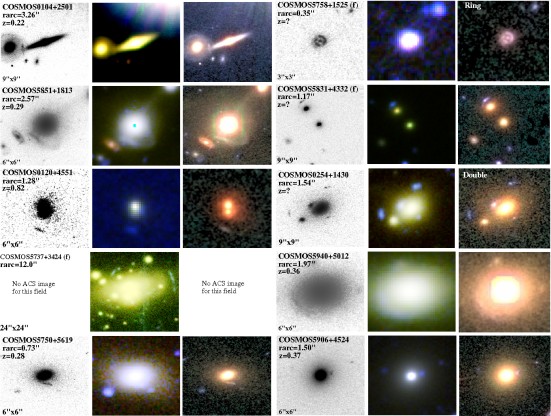}
\caption{\label{double} Other strong lensing candidates. For these
systems, no satisfying fit of the lensing galaxy was possible.
Therefore we display only the ACS I-band image of the system
(left panel when available),  the pseudo-color image (middle panel)
and the
color image sharpened with the ACS image (right panel). The labels refer to the
multiplicity of the images when different from single and when known.  The letter
(f) close to the name of an object means that it
was found fortuitously. North is to the top of the image and East is to the left.}
\end{center}
\end{figure*}

\begin{figure*}
\begin{center}
\includegraphics[width=14cm]{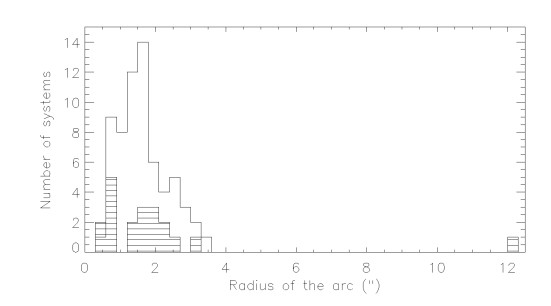}
\caption{\label{historarc}Distribution of the arc radii in the
 sample of COSMOS strong lens candidates (solid line histogram). The striped
histogram indicates the distribution of the best systems.  }
\end{center}
\end{figure*}

 The location of lens candidates in the absolute V-band magnitude 
 versus redshift plane are indicated by crosses in Fig.~\ref{mvz}.

  We divide the sample
in two sub-samples: the multiple arcs and long curved image systems, which we refer to hereafter as the
``best systems'', and less certain single arclet systems.
 Given their distinctive shape, the
 multiple images and long curved arcs have a greater probability of being
 genuine strong lenses relative to the single arclet systems. Throughout 
 the rest this paper we therefore restrict our study to the sample of ``best systems''.

The
  stamp images of the best systems are shown in Figs.~\ref{figure11},
\ref{figure22}, \ref{figure33} and \ref{double} and discussed in
Sect.~\ref{mas}.
  Images of the single arclet systems are displayed in
Figs.~\ref{toto1}, \ref{figure2},
  \ref{figure3}, \ref {figure4}, \ref{figure5}, \ref{figure6},
  \ref{figure7} and \ref{double} and discussed in Sect.~\ref{sas}. 
Their characteristics are  given in Tables
\ref{photom}, and \ref{photom1}.
In Fig.~\ref{historarc} we have reported the distribution of the arc
radius of the COSMOS strong lens candidates. Most systems have an arc radius$\lesssim$ 2\arcsec.

\subsection{The best systems}\label{mas}

\begin{figure}
\begin{center}
\includegraphics[width=7cm]{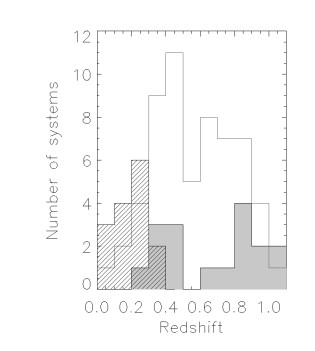}
\caption{\label{histoslacscastles} The spectroscopic redshift
distribution of the lensing galaxies in the SLACS survey (hashed
histogram) and in the CASTLES database (solid line histogram), compared
to the photometric redshift distribution of the best systems in the
COSMOS
survey (filled-in grey histogram). }
\end{center}
\end{figure}
In twenty cases, we have detected systems with a single long curved
image or with two or more arcs of similar color symmetrically located
around the lensing galaxy or with ring-like features.  Their
coordinates and characteristics are listed in Table \ref{photom}. We comment on
the most interesting or/and less trivial systems hereafter.

A bright arc-like feature is clearly visible in the ACS image of the
system 
COSMOS~0012+2015. The counter-arc becomes visible after subtraction of
the lensing galaxy profile, South-West of the center of the lens
galaxy.

The system COSMOS~0018+3845 is most probably a triply or possibly quadruply imaged
background galaxy. The North-West arc would be a single image of the
galaxy while the South-East arc would be a composite of two or three images
of the same galaxy.

The system COSMOS~0038+4133 is a complete Einstein ring
candidate, formed by the distorted image of a single background
galaxy.

Four systems show ring-like structures around a central elliptical
object (COSMOS~0049+5128, COSMOS~0124+5121, 
COSMOS~5941+3628 and COSMOS~5947+4752). In those cases, it is
difficult to conclude without any spectroscopic data whether the
ring is due to gravitational lensing or if it is a genuine part of the
central galaxy (like a ring-galaxy or a shell). The main argument in
favor of
them being lens candidates is the strong color difference
between the elliptical central objects and the ring features,
suggesting they
are not physically related.

There are probably four images of a background source in the
candidate system
COSMOS~0050+4901.  Three images are combined in an arc South-West
of the lensing
galaxy, while a counterpart image is visible North-East of the
galaxy.

The system COSMOS~0211+1139 displays two giant arcs, each with
different colors in the {\it Subaru} image. They are probably distorted
images of two different background galaxies.

The pseudo-color image of COSMOS0216+2955 shows a strong color
difference between the galaxy and the potential arcs. Nevertheless, the
galaxy subtraction shows features that are not fitted by the Sersic
profile. They seem to be linked to the galaxy core. Hence, there are
two possible explanations for this system: either it is actually a
strong lens, and the lensing galaxy is a spiral galaxy, or the two
arcs observed are part of the arms of a spiral galaxy.

The system COSMOS0254+1430 has  two lensing galaxies at very
short distance. We have not been able to make a clean galaxy light
profile
fit. The two arclets located North-East and South-West have are clearly
visible in the color image.

In the system COSMOS~5737+3424, we observe at least three giant
arc like features around the central galaxies of an identified XMM
detected
cluster (Finoguenov et al.~2007).

 COSMOS~5921+0638 is a quadruply-imaged lensed quasar or very
compact source. An Einstein ring is clearly visible in the ACS/F814W
band image.

\subsection{The single arclet systems }\label{sas}

\begin{deluxetable}{l l l l l l l l r l l l }
\tablewidth{0pt}
\tablecaption{\label{photom} Summary of the lensing galaxy parameters for the best systems}
\tablehead{
\colhead{Name} & 
\colhead{RA~~~~~~~~~~DEC} & 
\colhead{z$\pm\Delta$z} &  
\colhead{r$_{arc}$} & 
\colhead{R$_{eff}$} & 
\colhead{$n$}&  
\colhead{I$_{814w}^{l}$} & 
\colhead{PA} &
\colhead{$\epsilon$}  & 
\colhead{I$_{814w}^{s}$} \\
\colhead{ } & 
\colhead{h:m:s~~~~~~~~d:\arcmin:\arcsec} & 
\colhead{ } &
\colhead{\arcsec} & 
\colhead{\arcsec} & 
\colhead{ } &
\colhead{ } &
\colhead{deg} &
\colhead{ } & 
\colhead{ }
}
\startdata
0012+2015 &10:00:12.6 +02:20:15 &0.41$\pm^{0.03}_{0.05}$& 0.90
&0.53 &8.1&19.28$\pm^{0.01}_{0.01}$  &-10.2 & 0.25  &21.6  \\
0013+2249 &10:00:13.9 +02:22:49 &0.38$\pm^{0.06}_{0.02}$& 1.65
&0.90 &3.9 & 18.89$\pm^{0.01}_{0.01}$ & 38.2& 0.16  &22.4   \\
0018+3845$^f$ &10:00:18.4 +02:38:45 &0.71$\pm^{0.02}_{0.13}$& 0.40
&0.29 &5.6&23.60$\pm^{0.01}_{0.01}$ &-21.8 & 0.23  & 22.7 \\
0038+4133 &10:00:38.2 +02:41:33 &0.89$\pm^{0.05}_{0.03}$&  0.74
&0.72 &5.2&20.39$\pm^{0.05}_{0.05}$  & -2.9 &0.26
&  20.5 \\
0047+5023 &10:00:47.6 +01:50:23 &0.85$\pm^{0.07}_{0.05}$& 0.70
&0.72 &1.9 &20.65$\pm^{0.01}_{0.01}$& 33.4&0.05 &
22.8 \\
0049+5128 &10:00:49.2 +01:51:28 &0.33$\pm^{0.03}_{0.05}$& 2.22
&1.07 &1.2& 19.61$\pm^{0.01}_{0.01}$& -24.6&0.22
&23.3   \\
0050+4901$^f$ &10:00:50.6 +02:49:01 &1.01$\pm^{0.07}_{0.05}$&
1.90      &0.36 &2.9& 21.72$\pm^{0.02}_{0.03}$& 26.6 &0.39   &22.7 \\
0056+1226 &10:00:56.7 +02:12:26 &0.44$\pm^{0.04}_{0.04}$&  1.20
&0.96 &4.6 &18.70$\pm^{0.03}_{0.00}$ &40.3& 0.00
&23.3 \\
0124+5121 &10:01:24.5 +01:51:21 & 0.84$\pm^{0.04}_{0.04}$ &  0.84 &
0.24& 2.4&22.43$\pm^{0.15}_{0.05}$   &  -51.4&  0.25 &  23.2   \\
0211+1139 &10:02:11.2 +02:11:39 &0.90$\pm^{0.06}_{0.04}$&  3.20
&0.74 &4.1&21.09$\pm^{0.06}_{0.09}$ &11.0& 0.35 &
21.6  \\
0216+2955 &10:02:16.8 +02:29:55 &0.67$\pm^{0.05}_{0.03}$&  1.96
&0.90 &6.1 &19.98$\pm^{0.01}_{0.01}$ &-16.6 & 0.06  & 22.0 \\
0227+0451$^f$ &10:02:27.5 +02:04:51 &0.89$\pm^{0.03}_{0.05}$&
1.62  &0.30 &6.3 &21.94$\pm^{0.03}_{0.03}$   & 80.9 & 0.33 & 22.3 \\
0254+1430 & 10:02:54.0 +02:14:30 & \_& 1.54&\_ &\_ &\_ & -56.5 & 0.40
& 22.1\\
5737+3424$^{f}$ &09:57:37.0 +02:34:24 &  \_   & 12.0   & \_
&  \_  &\_ &    \_ &\_  &\_ \\
5758+1525$^{f}$& 09:57:58.6 +02:15:25 & \_     &  0.35   & \_   &
\_ & \_ & \_    & \_ &21.7\\
5857+5949 &09:58:57.0 +01:59:49 &0.39$\pm^{0.03}_{0.05}$&  2.15
&0.61 &1.4 & 20.05$\pm^{0.01}_{0.01}$   &-32.8& 0.36 & 21.9\\
5914+1219&09:59:14.7 +02:12:19 &1.05$\pm^{0.05}_{0.07}$ & 1.86
&0.27 &3.9 &23.25$\pm^{0.80}_{0.04}$     &14.4 &0.08  & 22.8 \\
5921+0638 &09:59:21.7 +02:06:38 &0.45$\pm^{0.03}_{0.05}$  & 0.80
&0.49 &6.8 &20.34$\pm^{0.02}_{0.02}$ &29.9& 0.08
&  20.6 \\
5941+3628 &09:59:41.3 +02:36:28 &0.90$\pm^{0.06}_{0.02}$  & 1.21
&0.76& 1.0&20.91$\pm^{0.02}_{0.02}$   & -5.4 & 0.08
  & 22.8\\
5947+4752 &09:59:47.8 +02:47:52 &0.28$\pm^{0.04}_{0.04}$  & 2.55&
0.51 & 1.6 &19.83$\pm^{0.04}_{0.05}$ & 4.1 &0.07& 22.8
\enddata
\tablecomments{
Column 1: Lens candidate  name.  A letter $^f$ indicates that the object was found fortuitously.
Columns 2 and 3: Coordinates in J2000.
Column 4: Photometric redshift of the lens and error bars (at 68\%
confidence level).
Column 5: Largest radius of the arc in arc-seconds.
Column 6: Effective radius of the lensing galaxy from the surface brightness fitting, in arcsec. 
Column 7: Sersic profile index.
Column 8: Total magnitude of the lensing galaxy in the WFPC/F814w band.
Column 9: Lensing galaxy position angle.
Column 10: Galaxy ellipticity $\epsilon$.
Column 11: Magnitude of the brightest image in the WFPC/F814w band (in
magnitudes per square arc-second).}
\end{deluxetable}

Forty seven of the discovered lens candidates have a single arclet
identified in both the ACS and the color images. In these cases, no
counter image for these arcs could be found in the residuals of the
galaxy-subtracted image. As observed on the {\it Subaru} data, the arclets
are generally much bluer than the lensing galaxies, and are likely to
be only weakly distorted by the galaxy. Their coordinates and
characteristics are give in Table
\ref{photom1}. We comment below on a few interesting
systems.

The system  COSMOS~0009+2455 shows two elongated features
symmetrically distributed around the elliptical galaxy. The two
features have slightly different colors, therefore if the system is a
genuine
lens,
the sources are probably two different objects.

The arclet in the system  COSMOS~0028+1919 is located West of the
lensing galaxy, and appears in blue in the color image.

In the system COSMOS~5805+0413, a giant blue arclet appears
between
two galaxies. The arc is slightly curved in direction of the
western galaxy but there are no counterpart images around any of the two
galaxies. The parameters of the early spectral type galaxy at the center of
the stamp image are presented in the Table~\ref{photom1}.


\begin{deluxetable}{l  l l l l l l r l l   }
\tablewidth{0pt}
\tablecaption{\label{photom1} Summary of the lensing galaxy parameters for the single arclet systems}
\rotate
\tablehead{
\colhead{Cosmos} & 
\colhead{RA~~~~~~~~~~DEC} & 
\colhead{z$\pm\Delta$z} &
\colhead{r$_{arc}$} &
\colhead{R$_{eff}$} & 
\colhead{$n$} & 
\colhead{I$_{814w}^{l}$} & 
\colhead{PA} &
\colhead{$\epsilon$} &
\colhead{I$_{814w}^{s}$}\\
\colhead{Name} & 
\colhead{h:m:s~~~~~~~~d:\arcmin:\arcsec} &
\colhead{ } &
\colhead{\arcsec} &
\colhead{\arcsec} &
\colhead{ } &
\colhead{ } &
\colhead{deg} &
\colhead{ } &
\colhead{ }
}
\startdata
0009+2455 &10:00:09.7 +02:24:55 &0.39$\pm^{0.04}_{0.04}$&2.63    
&0.70 &3.7 & 19.49$\pm^{0.01}_{0.01}$& -2.5& 0.31
&22.6 \\
0028+1919 &10:00:28.6  +02:19:19 &0.61$\pm^{0.03}_{0.05}$&1.87    
&0.45 &1.7 & 20.34$\pm^{0.01}_{0.03}$ & 34.9 & 0.13
& 22.6 \\
0029+4018 &10:00:29.6 +02:40:18 &0.83$\pm^{0.05}_{0.07}$&2.99    
&0.83  &2.6 &18.94$\pm^{0.02}_{0.02}$  &47.4& 0.00
& 22.7\\
0047+2931 &10:00:47.1 +02:29:31 &0.32$\pm^{0.04}_{0.04}$&2.00    
&0.73 &2.3 & 19.56$\pm^{0.01}_{0.01}$ &-38.4 & 0.29
 & 21.6  \\
0050+0357 &10:00:50.8 +02:03:57 &0.84$\pm^{0.04}_{0.04}$&0.66    
&0.31 &2.8 & 22.31$\pm^{0.02}_{0.02}$ &-44.7 & 0.40
 &  22.5     \\
0055+3821 &10:00:55.7 +01:38:21 &0.66$\pm^{0.06}_{0.06}$&1.05    
&0.40 &2.4 & 20.86$\pm^{0.01}_{0.01}$ &10.3  & 0.14
  & 21.8 \\
0056+2106 &10:00:56.7 +02:21:06 &0.65$\pm^{0.03}_{0.05}$&1.75    
&0.34 &3.4 & 21.21$\pm^{0.04}_{0.04}$ &84.7  & 0.02  & 23.3 \\
0104+2046 &10:01:04.9 +02:20:46 &0.94$\pm^{0.10}_{0.10}$& 3.54      
&0.11 &3.8& 24.27$\pm^{0.20}_{0.05}$    &-0.7& 0.31  &  21.9\\
0104+2501  &10:01:04.6 +02:25:01 &0.22$\pm^{0.06}_{0.02}$&
3.26    & \_   &  \_   & \_                          &  \_       &
\_  &  21.8\\
0105+4531 &10:01:05.3 +02:45:31 &0.44$\pm^{0.04}_{0.04}$& 1.00   
&0.47 &3.4 & 20.38$\pm^{0.01}_{0.01}$ &15.9  & 0.00    & 22.3\\
0107+0533 &10:01:07.9 +02:05:33 &0.90$\pm^{0.06}_{0.06}$& 1.71   
&0.24 &3.2 & 21.83$\pm^{0.01}_{0.07}$ &-70.4 & 0.04   & 22.7 \\
0108+5606 &10:01:08.0 +01:56:06 &0.97$\pm^{0.07}_{0.05}$& 1.25   
&0.20 &1.8 & 22.81$\pm^{0.02}_{0.02}$ &-61.5 & 0.24  & 23.4 \\
0120+4551&10:01:20.2 +01:45:51  &0.82$\pm^{0.06}_{0.06}$& 1.28
&  \_   &  \_   & \_                   &  \_       & \_
&   22.9\\
0148+2325 &10:01:48.1 +02:23:25 &0.40$\pm^{0.04}_{0.04}$& 1.43
&1.27 &2.9 &18.81$\pm^{0.01}_{0.01}$ &-35.1 & 0.00 &  22.0   \\
0208+1422 &10:02:08.5 +02:14:22 &0.41$\pm^{0.04}_{0.04}$& 1.59
&0.45 &7.7 &20.06$\pm^{0.07}_{0.05}$ &-52.0 & 0.00& 22.0    \\
0220+2335 &10:02:20.2 +02:23:35 &0.65$\pm^{0.03}_{0.05}$& 1.04
&0.60 &1.7 &20.99$\pm^{0.01}_{0.08}$ &87.2  & 0.00 &  21.8   \\
0221+3440 &10:02:21.1 +02:34:40 &0.41$\pm^{0.03}_{0.05}$& 1.57
&1.06 &5.5 &19.35$\pm^{0.01}_{0.01}$ &-22.7 & 0.04 &  22.3   \\
0236+4807 &10:02:36.0 +02:48:07 &0.34$\pm^{0.06}_{0.06}$& 1.27
&0.68 &1.8 &20.15$\pm^{0.01}_{0.01}$ &22.0  & 0.11& 23.2    \\
0248+1422 &10:02:48.4 +02:14:22 &0.60$\pm^{0.04}_{0.04}$& 1.38
&0.22 &5.5 &21.72$\pm^{0.11}_{0.05}$ &18.8  & 0.24& 22.8    \\
5748+5524 &09:57:48.0 +01:55:24 &0.93$\pm^{0.07}_{0.05}$& 2.21
&0.33 &1.4 &22.15$\pm^{0.02}_{0.02}$ &89.9  & 0.00& 21.6    \\
5750+5619 &09:57:50.7 +01:56:19 &0.28$\pm^{0.04}_{0.04}$&
0.73      &\_ &  \_   & \_                          &  \_       &
\_            &  21.8   \\
5752+2057 &09:57:52.1 +02:20:57 &0.44$\pm^{0.04}_{0.04}$& 0.98
&0.51 &1.0 &20.27$\pm^{0.01}_{0.01}$ &3.8   & 0.14&  21.6\\
5754+5952 &09:57:54.1 +01:59:52 &0.90$\pm^{0.06}_{0.02}$& 1.33
&0.22 &6.9 &22.23$\pm^{0.20}_{0.05}$ &-16.5 & 0.43& 22.6 \\
5805+0413 &09:58:05.6 +02:04:13 &0.88$\pm^{0.04}_{0.04}$& 1.61
&0.12 &4.8 &23.00$\pm^{0.04}_{0.02}$ &-81.6 & 0.16& 22.8 \\
5806+5809 &09:58:06.8 +01:58:09 &0.32$\pm^{0.04}_{0.04}$& 1.50
&0.43 &2.9 &19.14$\pm^{0.02}_{0.08}$ &-13.6 & 0.00& 21.7 \\
5821+4437$^f$ &09:58:21.4 +01:44:37 &0.67$\pm^{0.05}_{0.03}$&     
1.40  &0.21 &3.5 &21.84$\pm^{0.07}_{0.11}$ &00.0  &0.28& 22.9 \\
5829+3734$^f$ &09:58:29.9 +01:37:34 &0.51$\pm^{0.05}_{0.03}$&
2.97   &0.46 &1.6 &19.95$\pm^{0.00}_{0.03}$ &-0.1  &1.15 &22.3 \\
5831+4332$^f$ &09:58:31.0 +01:43:32 &             \_          &
1.17   &\_   &  \_   & \_                          &  \_       &
\_            & 21.0   \\
5841+4646 &09:58:41.4 +02:46:46 &0.36$\pm^{0.04}_{0.04}$& 0.93
&0.47 &0.9 &19.65$\pm^{0.01}_{0.01}$ & 21.1 & 0.12  & 21.0 \\
5851+1813 &09:58:51.9 +02:18:13 &0.29$\pm^{0.03}_{0.05}$&2.57 &
\_   &  \_   & \_                          &  \_       &
\_            &  21.9  \\
5856+4755 &09:58:56.1 +02:47:55 &0.32$\pm^{0.04}_{0.04}$& 2.97
&0.17 &1.1 &19.54$\pm^{0.01}_{0.01}$ &48.8  & 0.39
  & 22.9 \\
5906+4524 &09:59:06.5 +02:45:24 &0.37$\pm^{0.03}_{0.05}$&1.50
& \_   &  \_   & \_                          &  \_       &
\_            &  21.6  \\
5919+3853 &09:59:19.4 +01:38:53 &0.40$\pm^{0.04}_{0.04}$&
2.35      &0.26 &1.9 &20.59$\pm^{0.01}_{0.02}$ &-15.7 & 0.08 &  22.7 \\
5924+0852 &09:59:24.7 +02:08:52 &0.93$\pm^{0.07}_{0.05}$& 0.97
&0.17 &1.6 &23.09$\pm^{0.27}_{0.02}$ & -9.8 & 0.01  &22.7  \\
5925+3039 &09:59:25.8 +02:30:39 &0.91$\pm^{0.07}_{0.05}$&  0.98
&0.15 &1.2 &23.16$\pm^{0.05}_{0.04}$ & 89.9 & 0.38
& 23.2 \\
5929+1352 &09:59:29.9 +02:13:52 &0.36$\pm^{0.04}_{0.04}$& 2.46      & \_& \_& \ & \_& \_ & 22.0 \\
5929+4553 &09:59:29.0 +01:45:53 &0.54$\pm^{0.06}_{0.02}$&
1.56    &0.67 & 2.1& 19.99$\pm^{0.01}_{0.01}$&-23.2 & 0.06 & 22.3  \\
5931+0229 &09:59:31.1 +02:02:29 &0.32$\pm^{0.04}_{0.04}$&   1.48
&0.81 & 1.9& 19.20$\pm^{0.00}_{0.02}$&89.9 &0.01 &
21.6 \\
5932+1018 &09:59:32.0 +02:10:18 &0.92$\pm^{0.04}_{0.06}$&   0.68
&0.40 &1.2 & 21.92$\pm^{0.12}_{0.03}$&19.9 & 0.22  & 22.6\\
5936+3621 &09:59:36.7 +02:36:21 &0.25$\pm^{0.03}_{0.05}$&   2.40
&0.71 &0.9 &19.14$\pm^{0.02}_{0.01}$ &-42.4  & 0.15
& 22.1 \\
5939+3044 &09:59:39.1 +02:30:44 &0.73$\pm^{0.03}_{0.05}$&   0.70
&0.58 &1.7 & 20.08$\pm^{0.01}_{0.05}$&76.9  &0.00&
22.9 \\
5940+3253 &09:59:40.4 +02:32:53 &0.41$\pm^{0.03}_{0.05}$&   1.53
&0.51 &1.4 &19.45$\pm^{0.01}_{0.01}$ &58.4  & 0.40
& 22.5  \\
5940+5012 &09:59:40.0 +02:50:12&0.36$\pm^{0.04}_{0.04}$&1.97   &
\_   &  \_   & \_                          &  \_       &
\_            &  21.8  \\
5942+2829 &09:59:42.7 +02:28:29 &0.72$\pm^{0.04}_{0.04}$& 1.39
&0.31 &1.6 &21.88$\pm^{0.03}_{0.05}$ &34.0  &0.04
&  22.8 \\
5943+2816 &09:59:43.1 +02:28:16 &0.39$\pm^{0.05}_{0.03}$&
1.64         &0.69 &1.3 &19.43$\pm^{0.01}_{0.01}$ &79.9  &0.01  & 
22.4 \\
5951+1236 &09:59:51.0 +02:12:36 &0.46$\pm^{0.06}_{0.02}$&
1.53       &1.21 & 2.2&18.66$\pm^{0.01}_{0.01}$ &20.9  &0.29 & 21.9\\
5959+0348 &09:59:59.7 +02:03:48 &0.82$\pm^{0.06}_{0.06}$& 1.20
&0.20 &3.5 &21.62$\pm^{0.04}_{0.04}$ &-49.2 & 0.20
& 22.8 
\enddata
\tablecomments{Column 1: Lens candidate name.
Columns 2 and 3: Coordinates in J2000.
Column 4: Photometric redshift of the lens and error bars (at 68\%
confidence level).
Column 5: Radius of the arc in arcsec.
Column 6: Effective radius of the lensing galaxy from the galaxy fit
in arcsec.
Column 7: Sersic profile index.
Column 8: Total magnitude of the lensing galaxy in the WFC/F814w band.
Column 9: Lensing galaxy position angle.
Column 10: Galaxy ellipticity $\epsilon$=(a-b)/(a+b).
Column 11: Magnitude of the brightest image in the WFC/F814w band (in
mag~\arcsec$^{-2}$).
A letter $^f$ indicates that the object was found fortuitously.}
\end{deluxetable}

 \subsection{Radio and X-ray counterparts}
Six of the lens candidates have radio counterparts in the 1.4GHz
VLA-COSMOS data (Schinnerer et al.~2006). The radio data cover the
full COSMOS
field to a depth of $\sim$10.5$\mu$Jy$/beam$ rms in the center of the
field. All six sources lie above a 5$\sigma$ level in total flux,
ranging from
77\,$\mu$Jy to 1.24\,mJy in total 1.4GHz flux density. The distance
between
optical and radio position is in all cases less than 0.5\arcsec.

\begin{figure*}
\begin{center}
\includegraphics[width=14cm]{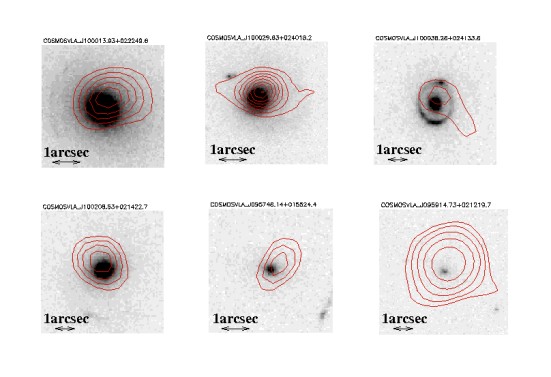}
\caption{\label{radiofig} Radio contours on top of the ACS images.
The name on top of each
  image refers to the radio source name in the VLA-COSMOS survey
(Schinnerer et
  al.~2007), and
 are based on the radio source exact coordinates. The contours are
defined as
 follows: 3, 5, 7, 9, 11, 13, 15 and 17$\sigma$ for COSMOS~0013+2249 and
0029+4018;
3, 4, 5 and 6$\sigma$ for COSMOS~0038+4133, 0208+1422 and 5748+5524;
3, 6, 12, 24 and 48$\sigma$ for COSMOS~5914+1219.}
\end{center}
\end{figure*}

One of the lens candidates, COSMOS 0013+2249, which is also a
radio
emitter, shows associated X-ray point source emission from the
XMM/Newton-COSMOS survey (Brusa et al.~2006). It has 0.5--2\,keV X-ray
emission of 3.74$\times$10$^{-15}$ $\pm$
0.34$\times$10$^{-16}$ erg$^{-1}$cm$^{-2}$s$^{-1}$, corresponding to
an X-ray luminosity of
1.21$\times$10$^{41}$ erg$^{-1}$s$^{-1}$ at the spectroscopic redshift of
the lensing galaxy ($z=0.347$, Trump et al.~2007). This X-ray emission as well as the
optical spectrum from Magellan (Brusa et al.~2006) are consistent with the lensing galaxy
being a radio galaxy with a central low-luminosity AGN.
In Table~\ref{radio} we list the radio counterpart properties and show
VLA overlays on the ACS images in Figure~\ref{radiofig}.
\begin{table*}
\renewcommand{\arraystretch}{1.0}
\centering
\begin{center}
\caption{\label{radio}  Radio information for the strong lenses.}
\begin{tabular}{l  l l l l  }
\hline
Name       & Flux & S/N & Flag & Comment\\
      &(mJy) &     &     &  \\
\hline
\hline
0029+4018  &       0.291& 16.710 &      1& likely
lensing galaxy counterpart\\   
0208+1422 &       0.088&  6.960 &      0&  likely lensing
galaxy counterpart\\   
5748+5524&       0.163&  5.680 &      0& likely lensing
galaxy counterpart\\
\hline
0013+2249 &       0.141& 12.450 &      1& likely lensing
galaxy counterpart\\
0038+4133 &       0.108&  4.740 &      0&     elongated,
likely lensing galaxy  counterpart \\
5914+1219 &       1.240& 88.520 &      1&     strong source,
must be a quasar in the galaxy source\\
& & &  & or in the lens\\
\hline
\end{tabular}
\end{center}
\tablecomments{Column 1: Lens candidate name.  Column 2: Total Flux in mJy.  Column
3: Signal to
noise ratio of the detection.  Column 4: Resolved flag (1=resolved
radio source, 0=point source).
Column 5: Comment.  }
\end{table*}

\section{Simple mass models for the best systems}\label{model}
 We want to estimate the Einstein radius of the lensing galaxies
and determine if there are multiple images which are not detected, and/or if there
is a need to take into account for the environment to model the system.
We use the Lenstool code to make a parametrised mass model of the lens
(Kneib et al.~1996, Jullo et al.~2007). The lensing galaxy is modeled by a
Singular Isothermal Ellipsoid (SIE).
An external shear contribution is added
when the SIE alone cannot reproduce the image configuration.
The parameters optimized by the code are the coordinates, the
orientation, the ellipticity and the velocity dispersion of the SIE,
and  the direction and amplitude
of the external shear. 
The (unknown) redshift of the source is calculated assuming that the source is at twice the comoving distance of the lens. This assumption, while clearly incorrect in general, 
 produces a lower limit for the estimated velocity dispersion of 
 the lensing potential, since the efficiency of lensing is greatest 
 for this configuration.

The Einstein radius is derived
using the
following equation:
\begin{equation}
{\rm R}_{E}=4\pi\left(\frac{\sigma_v}{c}\right)^2\frac{D_{ls}}{D_{s}}
\end{equation}
where $D_{s}$ is the angular distance between the observer and the
source and $D_{ls}$ is  the angular distance between the lensing
galaxy and the source.
The constraints used to optimize the mass model are the coordinates of the images. The multiplicity of
the
systems used in the model is given in Table \ref{modeltable}.
We first model the lens potential with a single SIE, and let one or more of the SIE parameters vary freely. The model uncertainties are
 typically $\pm$0.1 for the ellipticity, and $\pm$10\deg\, for the
orientation,
 reflecting a possible misalignment between the mass and the light
(Kochanek 2002).
If necessary, and if there are enough constraints, an external shear
contribution can be added.
For the ring-like objects, we have optimized the velocity
 dispersion of the SIE by hand in order to retrieve a perfect Einstein ring. For the purposes of this calculation, we
 assume that the source, the lens and the observer are perfectly
 aligned. Thus we have not given any $\chi^2$ value for these cases in Table \ref{modeltable}. 
 The system COSMOS~0013+2249 is not included in the table, as it possesses a
 single visible elongated and curved arc, providing too few constraints to fit a mass model.

In Table \ref{modeltable} we display the velocity dispersion and
the  external
shear parameters corresponding to the
best fits. Additional details are given in comments in the caption when
 the result of the fitting procedure is not straightforward.

\begin{table*}
\renewcommand{\arraystretch}{1.0}
\centering
\begin{center}
\caption{\label{modeltable} Morphological parameters and Einstein radii
of the lensing galaxies for the best systems.}
\begin{tabular}{c| c c |c|c c|l| c|c}
\hline
Name  & z$_l$&z$_s$        &$\chi^2$& $\sigma_v$    &
R$_{E}$& $\gamma$,$\theta_{\gamma}$ & More images? &Multiplicity   \\
      &  &     &   &(km~s$^{-1}$)   & (\arcsec)  &  &
Comments & \\
\hline
\hline

0012+2015 & 0.41&0.95 & 0.01&215.3  & 0.67   &no &no & Double\\
0018+3845 &0.71 &1.93 & 1.9 & 303.1   & 1.32    & (0.28,47.5)   &no  &Triple \\
0038+4133 & 0.89& 2.70& 0.08& 225.3 & 0.73  & (0.06,179.1) & no &Ring\\
0047+5023 &0.85 & 2.51& 1.4&313.0   & 1.41    & (0.23,174.1)& no &Triple\\
0049+5128 &0.33 & 0.74& \_  &380.0  & 2.09   &no &no (1,2)&Ring\\
0050+4901 &1.01 &3.34 & 3.4&342.3   & 1.69    &(0.23,76.1) & no (2,3)&Quad\\
0056+1226 &0.44 &1.03 & 1.3&337.4   & 1.64    & no& no& Double\\
0124+5121 &0.84 &2.47 & \_  &245.0  & 0.86   &no & no (1)& Ring\\
0211+1139 &0.90 &2.76 & 0.01&466.3  & 3.14   & no & no (2)& Double \\
0216+2955 & 0.67& 1.77& 0.06&348.5  & 1.75   & no&no (2) & Double\\
0227+0451 & 0.89& 2.70& 20.0&428.3  & 2.64   &no & yes (2,3)& Double\\
5857+5949 &0.39 &0.89 & 16.2&398.1 & 2.28  &no & yes (2,3)& Double\\
5914+1219 & 1.05&3.57 & 1.2&338.6  & 1.65  & (0.28,68.8)& yes (2) & Triple \\
5921+0638 & 0.45 &1.06& 0.48&221.0  & 0.70   & (0.09,100.5)& no &  Quad \\
5941+3628 & 0.90 &2.76& \_  &285.0  & 1.17   & no & no (1)& Ring\\
5947+4752 & 0.28 &0.61& \_  &370.0  & 1.97   & no & no (1)& Ring\\

\hline
\end{tabular}
\end{center}
\tablecomments{Column 1: name. 
Columns 2 and 3: lens and source redshift used for the modeling. 
Column 4: $\chi^2$ of the fit between the best mass model and the data. 
Columns 5 and 6: velocity dispersion and Einstein radius of the lens. 
Column 7: External shear contribution? if yes, value and orientation of the shear.
 Column 8: Do more images appear during the modeling? yes or no. The labels refer to the comments given hereafter.  
Column 9: Multiplicity of the images used for the mass modeling.   }
\tablecomments{
(1) The mass model is created using a source aligned with the lens center,
and we scale
$\sigma_v$ in order to match the ring size.
(2) The angular distance between the images and the velocity dispersion of the
lens makes us think  that
the lensing galaxy is associated with a galaxy group or cluster. 
(3) The $\chi^2$ value is high, suggesting that this system is not a genuine lens, or that a simple mass model is not a good representation of the total lens potential.
 }
\end{table*}


\section{Analysis of the sample and comparison to other strong
lensing surveys}\label{seccomparison}
In this section we compare the physical parameters of the foreground
in the COSMOS strong lens sample with independent galaxy-galaxy
strong lensing
catalogs from the SLACS survey (Treu et al.~2005, Bolton et al.~2006)
and the
CASTLES database\footnote {http://www.cfa.harvard.edu/castles/}
(Mu\~noz et
al.~1998).  Although the SLACS and CASTLES samples were
selected very differently from the COSMOS strong lensing sample, they also
have {\it HST} imaging. Thus a rough comparison in terms of the galaxy lensing properties is warranted, at least for the sub-sample of best systems.

\subsection{Redshift and effective radius}

\begin{figure}
\begin{center}
\includegraphics[width=7cm]{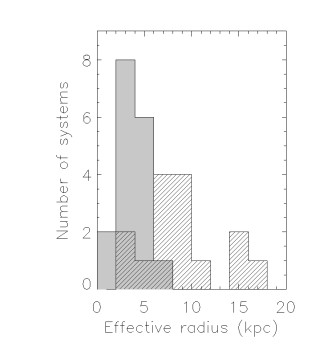}
\caption{\label{historeff} The effective radii of
strong lenses in the SLACS survey (hashed histogram) and the best
systems
from the COSMOS survey (histogram filled in grey).}
\end{center}
\end{figure}

Photometric redshifts have been measured by Mobasher et al.
(2007) for most of the COSMOS strong lensing galaxy candidates (see
Table \ref{photom1}, and \ref{photom}). We first
compare the redshift distribution of the  lensing galaxy candidates
found in the COSMOS field (17 best candidates with a photometric
redshift)  to the lensing galaxies in  the CASTLES database
(90 strong galaxy-galaxy lenses, lens redshift known for 59 of
them) and in the SLACS survey (19 strong galaxy-galaxy lenses) (Fig.
\ref{histoslacscastles}). The (photometric) redshift distribution of
the COSMOS sample of strong lenses extends to significantly higher
redshifts than the
spectroscopic redshift distribution of galaxies in the SLACS sample,
which is intrinsically limited due to the SDSS depth, but is similar
to the spectroscopic redshift distribution of galaxies in the CASTLES
sample.

The effective radii and arc radius of the lensing galaxies (in kpc)  are
simply obtained by converting the effective radii in angular units to
physical distances, via the photometric redshift of the lensing
galaxy. The SLACS sample includes lenses with larger effective lensing
radii than the COSMOS sample (see Fig. \ref{historeff}). This is probably an observational bias,
as (for practical reasons) we inspect visually the lens candidates in a
fixed 10\arcsec$\times$10\arcsec, box, and therefore miss any galaxies
that would generate arcs at larger radii.

\subsection{The absolute  magnitude}

The absolute magnitudes of the SLACS lensing galaxies are available in
Treu et al.~(2006). The mean value of their V-band magnitude is $\langle M_V\rangle=-22.71\pm^{\rm 1.23}_{\rm 2.46}$. The absolute V-band
magnitudes of the lensing galaxy candidates of our ``best systems'' sample have
a mean value of
$\langle M_V\rangle=-21.92\pm^ {\rm 1.78}_{\rm 1.23}$. 
Thus the
 COSMOS sample reaches slightly fainter magnitudes than the SLACS sample,
 despite the fact that it extends to higher lens redshifts.

\subsection{The velocity dispersion}
\begin{figure}
\begin{center}
\includegraphics[width=7cm]{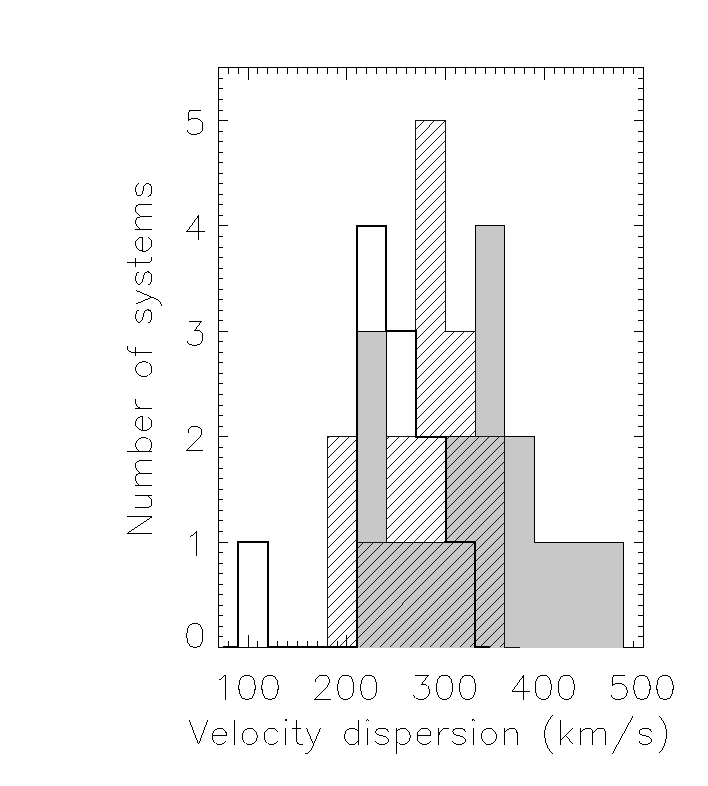}
\caption{\label{histosig} The velocity dispersions of
strong lenses in the SLACS survey (hashed histogram) and in the
CASTLES database (solid line) compared to the velocity dispersions of
the best systems in the COSMOS sample (histogram
filled in grey). }
\end{center}
\end{figure}

The velocity dispersions
 of the lensing galaxies of the COSMOS sample are derived from
simple mass models presented in Sect. \ref{model}. They  are displayed in Table
\ref{modeltable}. We have compared the  velocity dispersions of the lensing galaxy candidates of the best systems to the velocity dispersions of the lensing
galaxies in the SLACS and in the CASTLES samples (also measured using isothermal mass potentials, Fig. \ref{histosig}). The
lensing galaxies in the COSMOS sample have a slightly higher velocity dispersion than the lensing galaxies in SLACS and in CASTLES.
The difference with the SLACS sample can be due to the fact that we are probing in mean larger arc radius (therefore larger Einstein radius). 
 In addition, we have estimated the source redshift from the lensing galaxy photometric redshifts, that have an uncertainty. This can lead in some cases to wrong estimations of the velocity dispersion of the main deflector.  
In particular, we notice that two systems have a high velocity dispersion  ($>$
400~km~s$^{-1}$) :
COSMOS~0211+1139 and
COSMOS~5914+1219. 
 If real, these values suggest that these galaxies are in the center of a group, and that we are measuring here both the galaxy mass and the group mass.
  
\begin{figure}
\begin{center}
\includegraphics[width=7cm]{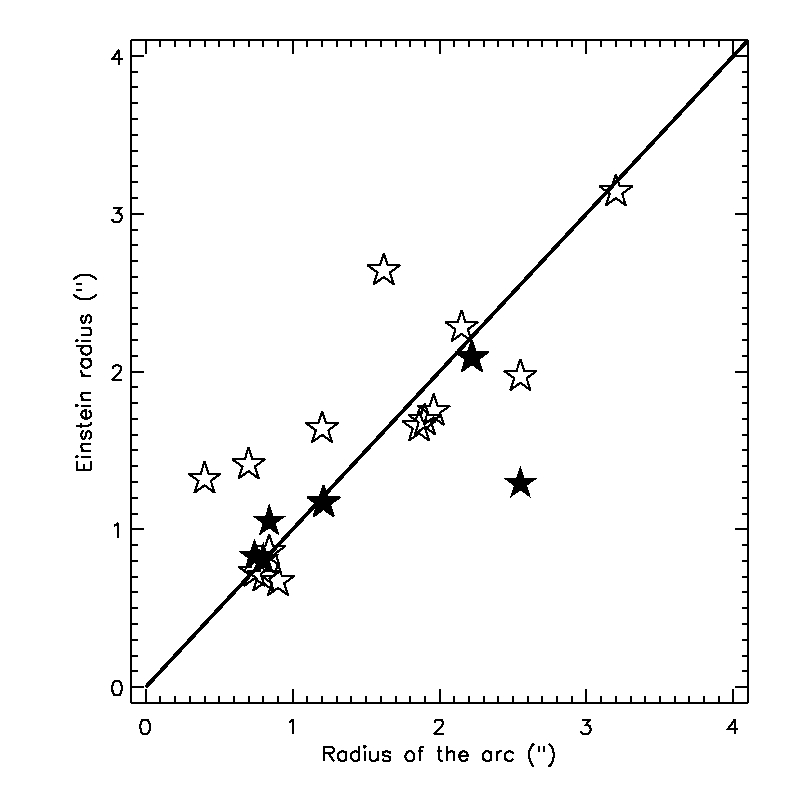}
\caption{\label{rarcrein} The arc radius versus the Einstein
radius of the best systems. The black stars show the ring like candidates.  The solid line traces
R$_E$=r$_{arc}$.}
\end{center}
\end{figure}
\subsection{The Einstein radius} \label{einsteinradius}



The Einstein radius is directly derived from the velocity
dispersion of the
lensing galaxy, it is therefore another mass estimator for the lens
potential. It has the advantage of being directly comparable to the
observable
arc radius.
  The value of the Einstein radii of the lensing galaxies R$_E$ are
given in Table~\ref{modeltable}.
 For the best systems,
the Einstein radius is expected to be  similar to the arc radius, while
an equality between these two values is expected in case of perfect
alignment between the observer, the source and the lens (Einstein
ring).
The plot in Fig.~\ref{rarcrein} shows  that the two radii are indeed similar for 
 most of the observed systems.
 Nevertheless we observe some exceptions: some systems have a large Einstein radius in comparison to their arc radius: COSMOS~0018+3845 ($r_{arc}$=0.40\arcsec, $R_{E}$=1.44\arcsec),
0047+5023 ($r_{arc}$=0.70\arcsec, $R_{E}$=1.29\arcsec) and 5914+1219
($r_{arc}$=1.86\arcsec, $R_{E}$=2.82\arcsec). 
For the three systems, there is a strong
external shear
added to the SIE. This may indicates the presence of a massive
structure in the direction to the source or in its
vicinity that could have stretched the angular separation between the images  and/or   biased strongly the estimation of the velocity
dispersion, and hence of the estimation of the  Einstein radius
of the lensing galaxy.

\section{Discussion and Conclusion}\label{secdiscussion}

In this paper we present 67 new strong galaxy-galaxy lensing
candidates found by visual inspection of the COSMOS field. From a
catalog of 278,526 galaxies with photometric redshifts (Capak et
al.~2007), we have selected a sub-catalog of 9452 bright galaxies
(M$_V<$-20~mag), at intermediate redshift ($0.2 \leq z_{phot} \leq 1.0
$) and spectrally early type. The selection criteria for this
sub-catalog were mainly motivated by the purpose of reducing the
sample of galaxies to be visually inspected. From the visual
inspection of
ACS/I$_{F814w}$-band stamp images of 10\arcsec$\times$10\arcsec, we
have built a sample of 337 possible lenses with sizes up to
r$_{arc}\sim$5\arcsec. This number is reduced to 67 strong
 galaxy-galaxy lensing candidates after a final cut based on an inspection of the
 pseudo-color images, and the fit and subtraction of the lensing galaxy
luminosity profile. There are 47 candidates with a single arclet and
20 candidates with multiple images of a background source or with a
large curved arc, called the ``best systems'' in this paper. We have
produced a simple mass model (SIE+shear) for each of the best candidates. In a
few cases, we have also been able to predict the existence and
direction of an
external shear contribution. Comparing the velocity dispersions
 to those of lenses in the the SLACS and CASTLES samples, it is
 apparent that we are probing a slightly higher range of velocity
 dispersion.
Nevertheless, massive objects in the  immediate vicinity of the
COSMOS lenses, and the rough assumption concerning the source redshift, 
can explain this result.
We have inferred the Einstein radius for the lensing galaxies
considering that they are isothermal spheres, and compared these values
with the arc radius. The derived Einstein radii agree well with the
 observed arc radii, making us confident that the
 sample consists mainly of genuine lens systems.


The ``visual inspection''  approach is long and tedious, and does not
ensure a
full understanding of the completeness of the strong lenses
sample. Nevertheless, our meticulously-obtained large sample of strong
lenses, even if incomplete, provides a valuable catalog to help
develop, test and improve automatic strong lens-finding algorithms for
similar data-sets. Such work is currently in progress (Cabanac et
al.~2007, Marshall et al.~2007, Seidel \& Bartlemann 2007).  Also, in future work (Faure et
al.~2007) the direct study of the spatial correlation between
large-scale structure in the COSMOS field (Scoville et al.~2007,
Massey et al.~2007) and the strong lenses discovered here will allow
us to measure the impact of the total, projected mass distribution
along the line of sight to a given strong lens, as recently
 investigated for three strong lenses in the AEGIS survey (Moustakas et
 al.~2006). A spectroscopic follow-up of the lens candidates is in
progress at the VLT using the FORS1-Multi-Object Spectroscopy
capability. This will allow us to measure precisely the redshift of
both the lens and the source, and to explore the close neighborhood of the lens, thus allowing a definite measurement of
the mass of each strong lensing galaxy.

Finally, this study gives a lower limit on the number of strong lenses expected in future
deep space surveys such as the proposed JDEM/SNAP multicolor weak-lensing survey. Indeed, we expect to find  at least 10 strong lenses per square degrees. Thus for a 10 000 square degree survey, we should discover about 100 000 strong lenses
thus allowing a high precision statistical analysis of the mass properties of galaxies and its evolution over time.
 The sample of strong lenses is available at this address: http://cosmosstronglensing.uni-hd.de/ 

\acknowledgments

The HST COSMOS Treasury program was supported through NASA grant
HST-GO-09822. We wish to thank Tony Roman, Denise Taylor, and David
Soderblom for their assistance in planning and scheduling of the
extensive COSMOS observations.  We gratefully acknowledge the
contributions of the entire COSMOS collaboration consisting of more
than 70 scientists.  More information on the COSMOS survey is
available at \url{http://www.astro.caltech.edu/cosmos}.
Based on observations with the NASA/ESA {\em Hubble Space Telescope},
obtained at the Space Telescope Science Institute, which is operated
by AURA Inc, under NASA contract NAS 5-26555; also based on data
collected at: the Subaru Telescope, which is operated by the National
Astronomical Observatory of Japan; the European Southern Observatory
under Large Program 175.A-0839, Chile; Kitt Peak National Observatory,
Cerro Tololo Inter-American Observatory, and the National Optical
Astronomy Observatory, which are operated by the Association of
Universities for Research in Astronomy, Inc. (AURA) under cooperative
agreement with the National Science Foundation; and the
Canada-France-Hawaii Telescope operated by the National Research
Council of Canada, the Centre National de la Recherche Scientifique de
France and the University of Hawaii.  It is a pleasure to acknowledge
the excellent services provided by the NASA IPAC/IRSA staff (Anastasia
Laity, Anastasia Alexov, Bruce Berriman and John Good) in providing
online archive and server capabilities for the COSMOS datasets. CF thanks the
support from the European Community's Sixth Framework Marie Curie
Research Training Network Programme, Contract No.  MRTN-CT-2004-505183
``ANGLES''.  JPK thanks support from CNRS and Caltech. GC is partially
supported by the Cordis ERG proposal n. 029159. We acknowledge the referee for their useful remarks and J. Duke for correcting the draft.
  
{\it Facilities:}
\facility{HST (ACS)},
\facility{Subaru (Suprime)}.
\facility{CFHT (Megacam)}.




\end{document}